\documentclass[aps,pre,twocolumn]{revtex4}
\usepackage[titletoc,toc,title]{appendix}
\usepackage{amsmath}
\usepackage{amssymb}
\usepackage{natbib}
\usepackage{color}
\usepackage{graphicx}
\usepackage{subfigure}

\begin{document}

\bibliographystyle{apsrev}

\title{Stiffer double-stranded DNA in two-dimensional confinement due to bending anisotropy}

\author{H. Salari}
\address{Department of Physics, Sharif University of Technology, P.O. Box 11155-9161, Tehran, Iran.}
\author{B. Eslami-Mossallam}
\address{Department of Bionanoscience, 
Kavli Institute of Nanoscience,
Delft University of Technology, 
Lorentzweg 1, 2628 CJ 
Delft, The Netherlands.}

\author{H. F. Ranjbar}
\address{Institute of Complex Systems (ICS-2), Forschungszentrum Jülich, Wilhelm-Johnen-Straße, 52425 Jülich, Germany.}

\author{M. R. Ejtehadi}
\email{ejtehadi@sharif.edu}
\address{Department of Physics, Sharif University of Technology, P.O. Box 11155-9161, Tehran, Iran.}
\address{School of Nano Science, Institute for Research in Fundamental Sciences (IPM), Tehran 19395-5531, Iran.}

\begin{abstract}
Using analytical approach and Monte-Carlo (MC) simulations, we study the elastic behaviour of the intrinsically twisted elastic ribbons with bending anisotropy, such as double-stranded DNA (dsDNA), in two-dimensional (2D) confinement. We show that, due to the bending anisotropy, the persistence length of dsDNA in 2D conformations is always greater than 3D conformations. This result is in consistence with the measured values for DNA persistence length in 2D and 3D in equal biological conditions. We also show that in 2D, an anisotropic, intrinsically twisted polymer exhibits an implicit twist-bend coupling, which leads to the kink formations with a half helical turn periodicity along the bent polymer.
\end{abstract}

\maketitle
\section{Introduction}
\begin{table}[h]
\caption{Some reported values of the persistence length of double-stranded DNA measured by various techniques in 3D conformations.\label{table:1}}
\begin{tabular}{p{4.5cm}ccc}
\hline
\hline Buffer composition & Method & $P_{\mathrm{(3D)}}$ (nm) & Ref. \\
\hline 
10 mM $\mathrm{Na}^+$, pH 7.0, $25^{\circ}\mathrm{C}$ & DE\footnote{DNA Stretching} & $47.4 \pm 1.0$ & \cite{Wang1997a} \\
93.0 mM $\mathrm{Na}^+$, pH 7.0, $25^{\circ}\mathrm{C}$ & DE & $43.8 \pm 1.4$ & \cite{Baumann1997} \\
10 mM $\mathrm{Na}^+$ & DE & $50$ & \cite{Smith1992} \\
200 mM $\mathrm{K}^+$, 10 mM Tris-HC1,  pH 7.2, $25^{\circ}\mathrm{C}$ & LS\footnote{Light Scattering} & $48 \pm 1$ & \cite{Nordmeier1992} \\
100 mM $\mathrm{Na}^+$, $20^{\circ}\mathrm{C}$ & LS & $45$ & \cite{Kam1981} \\
110 mM $\mathrm{Na}^+$, pH 7.4, $25^{\circ}\mathrm{C}$ & TPM\footnote{Tethered Particle Motion}  & $47.8 \pm 0.7$ & \cite{Brunet2015} \\
100 mM $\mathrm{Na}^+$, $20^{\circ}\mathrm{C}$ & FD\footnote{Flow Dichroism} & $48$ & \cite{Rizzo1981}\\
101 mM $\mathrm{Na}^+$, $20^{\circ}\mathrm{C}$ & TED\footnote{Transient Electric Dichroism} & $44$ & \cite{Porschke1991} \\
0-162 mM $\mathrm{Na}^+$, 1 mM $\mathrm{Mg}^{2+}$, pH 7.8, $16^{\circ}\mathrm{C}$ & DC\footnote{DNA Cyclization} & $45.0 \pm 1.5$ & \cite{Taylor1990}  \\
89 mM Tris borate/2 mM EDTA, pH 8.3 & DC & 48.5 & \cite{Geggier2010}\\
moderate salt buffer & Cryo-EM\footnote{Cryo-electron microscopy} & $45$ & \cite{Bednar1995}\\
\hline 
\end{tabular} 
\end{table}
The bending flexibility of double-stranded DNA plays a crucial role in its interactions with other macromolecules, e.g., proteins. The most convenient measure of bending flexibility of a polymer is the ``persistence length" ($P$), which is defined as the correlation length of the tangent unit vector along the contour length~\cite{Benetatos2003}. Many experimental and simulational techniques have been performed to measure this quantity of the DNA molecule~\cite{Porschke1986,Taylor1990,Porschke1991,Baumann1997,Wang1997a,Lu2001,Brunet2015,Rivetti1996,Wiggins01,Mazur2014b,Podesta2005,Moukhtar2007,Kundukad2014,Japaridze2015,Cassina2016}, and characterize its dependence on the ionic strength~\cite{Porschke1991,Baumann1997,Podesta2005,Drozdetski2016}, temperature~\cite{Lu2001}, sequence~\cite{Geggier2010} and length scale~\cite{Noy2012,Fathizadeh2012,Hossein2015}. In the single molecule stretching experiment,  Baumann et al. have shown that the persistence length of a random DNA sequence in moderate salt buffer is around $45-50\,\mathrm{nm}$~\cite{Baumann1997}. Other bulk experiments, such as DNA-cyclization~\cite{Taylor1990,Geggier2010} and gel electro-phoretic mobility~\cite{Rizzo1981,Porschke1991}, also result in a value in this range (Table \ref{table:1}).

\begin{figure}[h]
\includegraphics[scale=0.45]{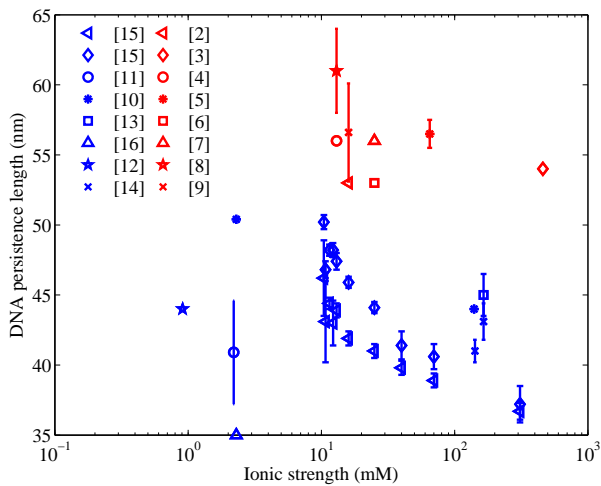}
\caption{Comparison between the reported values of $P_{\mathrm{2D}}$ (red dots) and $P_{\mathrm{3D}}$ (blue dots) for different ionic strength in the presence of divalent counter-ions. The ionic strength is defined as Ref.~\cite{Brunet2015}, and different markers correspond to different references as indicated in the legend.}
\label{fig:1}
\end{figure}

On the other hand, single-molecule imaging techniques, including Atomic Force Microscopy (AFM) and Electron Microscopy (EM), introduce an important new class of experiments to measure the persistence length of DNA molecule. In these experiments the molecules are attracted onto the surface of a substrate by divalent counter-ions, e.g. $\mathrm{Mg}^{2+}$~\cite{Rivetti1996,Pastre2003}. These divalent ions allow the molecule freely equilibrate in 2D and decrease the effects of the substrate on the chain statistics~\cite{Kundukad2014}. It is also known that, a small amount of $\mathrm{Mg}^{2+}$ in solution can dramatically decrease the persistence length~\cite{Baumann1997,Brunet2015}. Baumann et al. show that, with $0.1\, \mathrm{mM}$ of $\mathrm{Mg}^{2+}$, even with a small amount of monovalent counter-ions (e.g. 1.86 mM $\mathrm{Na}^+$), the persistence length of DNA decreases to $40.9 \pm 3.7\,\mathrm{nm}$~\cite{Baumann1997}. However, the DNA persistence length measured in 2D, $P_{\mathrm{2D}}$,
~\cite{Rivetti1996,Wiggins01,Mazur2014b,Podesta2005,
Moukhtar2007,Kundukad2014,Japaridze2015,Cassina2016}
 is generally bigger than the 3D values, $P_{\mathrm{3D}}$,~\cite{Porschke1986,Taylor1990,Porschke1991,Baumann1997,
 Wang1997a,Lu2001,Brunet2015} in the presence of divalent counter-ions (see fig~\ref{fig:1}). There are two experiments which do not address this discrepancy between 2D and 3D persistence lengths of dsDNA~\cite{Abels2005,VanNoort2004}. In these works different buffers were used in the 2D and 3D experiments, where $\mathrm{Mg}^{2+}$ is only present in the 2D experiment buffer. Thus, they are not shown in fig \ref{fig:1}. According to fig~\ref{fig:1}, the average value of $P_{\mathrm{2D}}$ is about $55.8\pm3.5\, \mathrm{nm}$ and remains almost constant with increasing the ionic strengths in contrast to $P_{\mathrm{3D}}$, which drops slowly.

This visible difference may arise from different effects of the divalent counter-ions within the experiments, i.e. measuring the persistence length in 2D and 3D conformations. It is known that in 2D, the divalent counter-ions bridge the negative charges of the phosphate backbone to the negatively charged mica surface~\cite{Rivetti1996,Abels2005}, while they act as intramolecular bridges between two phosphates in 3D conformations~\cite{Mantelli2011}. The bridges in the latter can greatly reduce the entropy of the chain and lead to low persistence lengths. In addition, the excluded volume interactions in 2D conformations can swell the molecule and therefore increase the persistence length~\cite{Rivetti1996,Hsu2012,Drube2010}. Rivetti et al. show that, although these interactions can increase the 2D persistence length, but this effect is negligible when DNA length is less than $1000\,\mathrm{nm}$ ($\simeq 3400 \,\mathrm{bp}$)~\cite{Rivetti1996}. But the molecules in 2D remain stiffer, even for lengths shorter than 3400 bp~\cite{Rivetti1996,Wiggins01,Podesta2005,Moukhtar2007,Kundukad2014,Japaridze2015}. Moreover, transition from B-DNA to A-DNA during the imaging in the dry air can also cause DNA stiffer~\cite{Rivetti2001,Hansma1996,Japaridze2016,Waters2016a}. To avoid this structural transition, DNA molecules in fig \ref{fig:1} were scanned in solution and biological conditions. Also, the errors in contour length estimation can affect the measured values for the persistence length. Underestimating by $\sim2\%$ leads to about $\sim60\%$ overestimation of $P_{\mathrm{2D}}$ for a 100 bp DNA~\cite{Wang2015}. But, this effect is reduced by increasing the contour length, where the overestimation decreases to $\sim10\%$ for a 500 bp DNA~\cite{Wang2015}. Finally, it has been also mentioned that the surface charges can affect the flexibility of DNA~\cite{Podesta2005}. Apart from above possibilities here we show that the anisotropic bending tendency of double stranded DNA increases the stiffness of the molecule in 2D.

The anisotropic bending of double-stranded DNA is a property of the sugar-phosphate backbone structure, in the sense that bending toward the grooves direction (roll) is much easier than toward the backbone direction (tilt)~\cite{Ulyanov1984}. Fourier analysis of free energy of DNA loops with lengths between $60-100$ bp shows two main oscillatory components~\cite{Saiz2005}. One with a helical period ($\sim 10.5$ bp) and another with a half helical period ($\sim 5.6$ bp) which may reflect the bending anisotropy. The sequence-dependent bending anisotropy of B-DNA has been observed in X-ray crystallography of DNA-protein complexes~\cite{Olson1998,Richmond} and NMR spectroscopy~\cite{Lankas2010}, as well as all-atomistic simulations~\cite{Teng2015,Lankas02,Bishop2005,Perez2008,Lavery2009,Lankas2006,Drsata2015,Ma2016}. Many theoretical studies have considered such anisotropic bending into the elastic models~\cite{Zhurkin1979,Balaeff1999,Norouzi2008,Eslami-Mossallam2008,Alim2007,Cheng2016}, and it is shown that although the bending anisotropy affect the elastic properties of a short DNA molecule in 3D, it becomes unimportant when the DNA segment is long enough to include a few full helical turns~\cite{Eslami-Mossallam2008,Becker2007}.

Here, we exploit methods from the statistical field theory as well as Monte-Carlo simulation technique to study the elastic properties of an intrinsically twisted ribbon with anisotropic bending in 2D. We show that it is possible to assign an effective persistence length to a long DNA molecule in 2D, similar to the 3D case. Whereas the isotropic bending model predicts equal persistence lengths in 2D and 3D, we show that due to the anisotropic bending the 2D persistence length is always bigger than the one in 3D. The difference between 2D and 3D persistence lengths depends on the relative strengths of the bending elastic constants (the strength of anisotropy) and also twist rigidity, while the latter implies an implicit twist-bend coupling in the model. The prediction of our model for the DNA persistence length in 2D is in good agreement with the experimental data, shown in fig \ref{fig:1}. Our finding can be relevant to other anisotropic chain polymers, e.g. double-stranded RNA or carbon nano-ribbons~\cite{Giusti2015}.  

\section{MODEL AND MATERIALS}
\subsection{The Planar Anisotropic Elastic Rod Model}
\begin{figure}[h]
\includegraphics[scale=0.25]{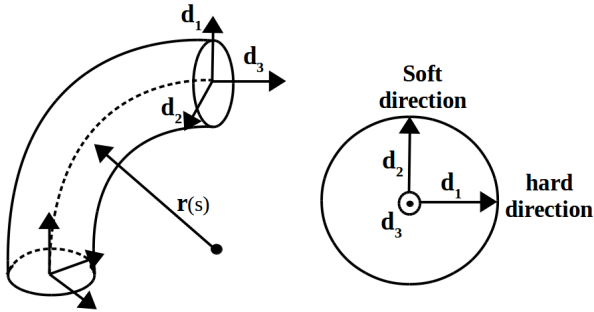}
\caption{Parametrization of the elastic rod.}
\label{fig:2}
\end{figure}

Double-stranded DNA is a helical nano-ribbon polymer which is represented as an anisotropic elastic rod. As fig \ref{fig:2} shows, at each point of arc length parameter $s$ on the centerline, $\vec{r}(s)$, one can attach an orthonormal basis $\{\hat{d}_1,\hat{d}_2,\hat{d}_3\}$, so called a ``material frame". There is two usual definitions for $\hat{d}_3$~\cite{Fathizadeh2012}, but in the simplest way it can be chosen to be along the tangent to the rod at every point so that $\hat{d}_3(s)=\hat{t}(s)\equiv(d/ds)\vec{r}(s)$~\cite{Heussinger2007}. $\hat{d}_1$ is along the grooves direction and points toward the major groove, and $\hat{d}_2=\hat{d}_3 \times \hat{d}_1$. Here we assume, bending about the $\hat{d}_2$-axis is easer than the $\hat{d}_1$-axis, therefore, the $\hat{d}_1$- and $\hat{d}_2$-axes correspond to the hard and soft bending directions (see fig \ref{fig:2}).

The derivatives of the orthonormal triads with respect to $s$ are defined as the following
\begin{equation}
\label{angularveloc}
\dot{\hat{d}}_i\equiv\frac{d}{ds}\hat{d}_i = \bold{\Omega}\times\hat{d}_i,\,\,\,\,\,\,\,i=1,2,3,
\end{equation}
where $\bold{\Omega}(s)=\kappa_1\hat{d}_1+\kappa_2\hat{d}_2+\omega\hat{d}_3$ is called the strain vector. The components of $\bold{\Omega}$ (i.e. $\kappa_1,\kappa_2$, and $\omega$) respectively correspond to rotation of the filament around $\hat{d}_1$, $\hat{d}_2$, and $\hat{d}_3$ and called tilt, roll and twist~\cite{Hossein2015}. Therefore, the elastic energy of an inextensible, unshearable and anisotropic filament in harmonic approximation can be written as~\cite{Eslami-Mossallam2008,Farshid2008}
\begin{equation}
\label{elasticener}
E/k_{\mathrm{B}}T = 1/2 \int_0^L ds(A_1\kappa_1^2+A_2\kappa_2^2+C(\omega - \omega_0)^2),
\end{equation}
where $A_1$ and $A_2$ are bending rigidities respectively for the hard and soft directions, $C$ is the twist rigidity and $\omega_0=1.8\, \mathrm{nm}^{-1}$ is intrinsic twist of B-DNA. In the elastic energy of eq. (\ref{elasticener}), we have ignored the explicit twist-bend coupling~\cite{Marko03,Liebl2015,Nomidis2016}.

For a planar DNA (confined in the $x$-$y$ plane where $\hat{z}.\hat{d}_3=0$), it is convenient to express the local triads in terms of Euler angles $\bold{\Theta}(s)=[\alpha(s),\beta(s),\psi(s)]$ ($0<\alpha<2\pi$, $\beta(s)=\pi/2$, $0<\psi<2\pi$) as the following~\cite{Eslami-Mossallam2008}:
\begin{eqnarray}
\label{dcomponents}
\hat{d}_1(s) &=& -\sin\psi\sin\alpha\,\hat{x}+\sin\psi\cos\alpha\,\hat{y}+\cos\psi\,\hat{z},\nonumber\\
\hat{d}_2(s) &=& -\cos\psi\sin\alpha\,\hat{x}+\cos\psi\cos\alpha\,\hat{y}-\sin\psi\,\hat{z},\\
\hat{d}_3(s) &=& \cos\alpha\,\hat{x}+\sin\alpha\,\hat{y},\nonumber
\end{eqnarray} 
where $\alpha(s)$ and $\psi(s)$ are respectively correspond to the local bend and twist angles. By substituting the eq. (\ref{dcomponents}) into eq. (\ref{angularveloc}) one can obtain the components of $\bold{\Omega}$,
\begin{eqnarray}
\label{omegacomponents}
\kappa_1(s) &=& \dot{\alpha}(s)\cos\psi(s),\nonumber\\
\kappa_2(s) &=& \dot{\alpha}(s)\sin\psi(s),\\
\omega(s) &=& \dot{\psi}(s).\nonumber
\end{eqnarray} 
and the local curvature is $\kappa(s)=\sqrt{\kappa_1^2+\kappa_2^2}=\dot{\alpha}$.

Experimentally, the persistence length of DNA in 2D conformation can be determined by measuring various statistical properties, such as the orientational correlation function~\cite{Faas2009,Abels2005}, the probability distribution of the bending angle~\cite{Wiggins01,Mazur2014b}, the mean-square end-to-end distance~\cite{Rivetti1996,Pastre2003,Moukhtar2007,Mantelli2011}, and force-extension~\cite{Maier2007}. The isotropic wormlike chain (WLC) model, i.e. $A_1=A_2=P$, has been widely used to fit the experimental data and obtain the persistence length. Below we derive each of these statistical properties for an anisotropic elastic model, where $A_1>A_2$.

For a chain with length $L$ and global bend angle $\theta(L)$ ($=\int_0^L\kappa(s)ds$) in 2D the free energy is given by the canonical relation:~\cite{Curuksu,Wiggins2005}
\begin{eqnarray}
\label{freeenergy}
F(\theta(L)) = -k_{\mathrm{B}}T\cdot\mathrm{ln}\left(p_{\mathrm{2D}}(\theta(L))\right),
\end{eqnarray}
where $p(\theta(L))_{\mathrm{2D}}$ is the normalised probability distribution of $\theta$ for 2D conformation. This probability distribution can be written as a path integral
\begin{eqnarray}
\label{npdb}
p_{\mathrm{2D}}(\theta(L)) &=& \mathcal{N}\int \mathcal{D}[\bold{\Theta}]\exp\left(-\int_0^L\frac{e[\bold{\Theta}(s)]}{k_{\mathrm{B}}T}ds\right)\times\nonumber\\
&&\delta\left(\theta-\int_0^L\kappa(s)ds\right),
\end{eqnarray}
where $\mathcal{N}$ is normalization constant and
\begin{equation}
\label{densener}
\frac{e[\bold{\Theta}(s)]}{k_{\mathrm{B}}T}=\frac{1}{2}(A_1\cos^2\psi+A_2\sin^2\psi){\dot{\alpha}}^2+\frac{1}{2}C(\dot{\psi}-\omega_0)^2
\end{equation}
is the density of elastic energy. In the case of the isotropic WLC model, $p_{\mathrm{2D}}(\theta(L))$ follows a Gaussian distribution as $\sqrt{2P/\pi L}\exp(-P\theta^2/2L)$, but in general, due to the first term of the right hand side of eq. (\ref{densener}), it is not easy to find an expression for $p_{\mathrm{2D}}(\theta(L))$. Using the probability distribution (\ref{npdb}), the tangent-tangent correlation function, $\langle\hat{d}_3(L).\hat{d}_3(0)\rangle_{\mathrm{2D}}=\langle\cos(\theta(L))\rangle_{\mathrm{2D}}$, as a function of $L$ is defined as~\cite{Rivetti1996}
\begin{eqnarray}
\label{correlation}
\langle\cos(\theta(L))\rangle_{\mathrm{2D}} \equiv \int_0^{+\infty}d\theta \cos(\theta(L)) p_{(\mathrm{2D})}(\theta(L)),
\end{eqnarray}
where, we are supposed $\theta\geq0$. It can be easily shown that for the isotropic case we have $\langle\cos(\theta(L))\rangle_{\mathrm{2D}}=\exp(-L/2P)$. Finally, the mean-squared end-to-end distance of the chain for short lengths, where the excluded volume interactions are negligible~\cite{Hsu2012}, is written by
\begin{eqnarray}
\label{endtoendequ}
\langle R^2\rangle_{\mathrm{2D}} &=& \int_0^L\int_0^L\langle\hat{d}_3(s).\hat{d}_3(s^\prime)\rangle_{\mathrm{2D}}dsds^\prime\nonumber\\
&=&\int_0^L\int_0^L\langle\cos(\theta(s-s^\prime)\rangle_{\mathrm{2D}}dsds^\prime,
\end{eqnarray}
where $\theta(s-s^\prime)$ is the angle between tangent vectors at points $s$ and $s^\prime$ along the contour. The above equation for the isotropic model can be directly determined by substituting $\langle\cos(\theta(s-s^\prime))\rangle_{\mathrm{2D}}=\exp(-(s-s^\prime)/2P)$ and straightforward integration as~\cite{Zoli2016}
\begin{eqnarray}
\label{endtoend2d}
\langle R^2\rangle_{\mathrm{2D}}=4PL-8P^2\left(1-\exp(\frac{-L}{2P})\right).
\end{eqnarray}

In case of an anisotropic elastic model the first term of the right hand side of eq. (\ref{densener}) implies an implicit twist-bending coupling for the model. In order to find the effects of this coupling, we evaluate $p_{\mathrm{2D}}(\theta(L))$ in two extreme limits: the large and small twist rigidity (i.e. $C\rightarrow\infty$ and $C\rightarrow 0$, respectively).
 
\begin{figure}[h]
\includegraphics[scale=0.45]{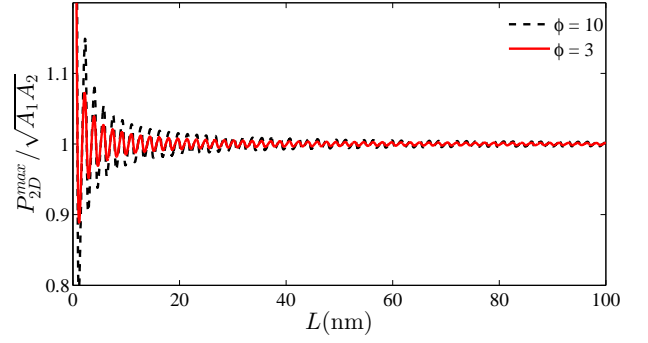}
\caption{The ratio of $P_{\mathrm{2D}}^{max}/\sqrt{A_1A_2}$ (eq. (\ref{P2dmax})) as a function of length for two different values of $\phi$.}
\label{fig:3}
\end{figure}

For large twist rigidity, the relative variations of $\dot{\psi}$ is negligible and $\dot{\psi}\simeq\omega_0$ which gives $\psi = \omega_0s$ (it is assumed the initial twist angle is zero). Therefore, the density of energy (\ref{densener}) reduces to 
\begin{eqnarray}
\label{densitymax}
e_{\infty}[\kappa(s)]/k_{\mathrm{B}}T=\frac{1}{2}\tilde{A}(\omega_0 s)\kappa(s)^2,
\end{eqnarray} 
where $\tilde{A}(\omega_0 s) = A_1\cos^2(\omega_0 s)+A_2\sin^2(\omega_0 s)$. By substituting this energy density into Eq. (\ref{npdb}) and replacing the Dirac delta function by the appropriate Fourier transform, one can find the probability distribution of $p_{\mathrm{2D}}(\theta(L))$, as follows
\begin{eqnarray}
\label{npdbmax}
p_{\mathrm{2D}}(\theta(L)) &&= \mathcal{N}\int_{-\infty}^{+\infty}\frac{dK}{2\pi}e^{iK\theta}\times\\
&&\int \mathcal{D}[\kappa]\exp\left(-\int_0^L(\frac{\tilde{A}(\omega_0 s)}{2}\kappa^2 +iK\kappa)ds\right).\nonumber
\end{eqnarray}
Then, upon a straightforward integration, this equation yields a Gaussian distribution
\begin{eqnarray}
\label{npdbmax2}
p_{\mathrm{2D}}(\theta(L)) =\sqrt{\frac{2P_{\mathrm{2D}}^{max}}{\pi L}}\times\exp\left(-\frac{P_{\mathrm{2D}}^{max}\theta^2}{2L}\right),
\end{eqnarray}
with 
\begin{eqnarray}
\label{P2dmax}
\frac{1}{P_{\mathrm{2D}}^{max}} &&= \frac{1}{L}\int_0^L\frac{ds}{\tilde{A}(\omega_0 s)} =\frac{1}{L\omega_0\sqrt{A_1A_2}}\times\\
&&\left(\tan^{-1}\left(\phi^{-1/2}\tan(L\omega_0)\right)+\pi\left[\frac{1}{2}+\frac{L\omega_0}{\pi}\right]\right)\nonumber,
\end{eqnarray}
where $\phi = A_1/A_2$. The bracket means ``integer part" which is
added to get rid of discontinuity in the $\tan^{-1}$ function. Fig \ref{fig:3} indicates the length dependent of $P_{\mathrm{2D}}^{max}/\sqrt{A_1A_2}$ for two different values of $\phi$. It shows  that, by increasing the chain length $P_{\mathrm{2D}}^{max}$ approaches soon to its asymptotic value $\sqrt{A_1A_2}$, then 
\begin{eqnarray}
\label{P2dmax_ext}
P_{\mathrm{2D}}^{max} = \sqrt{A_1A_2} \,\,\,\,\,\,\, \mathrm{for}\,\,\,L\omega_0\gg1,
\end{eqnarray}
which implies that at large twist rigidity limit, a long enough 2D anisotropic DNA behaves like an isotropic DNA with the bending constant $\sqrt{A_1A_2}$. 

On the other hand, in the limiting case of the small twist rigidity, there is no constrain on the twist degree of freedom, then the local twist angle, $\psi(s)$, is free to choose any value in the range of $[0,2\pi]$. Therefore, the energy density of (\ref{densener}) in this limit is rewritten as
\begin{eqnarray}
\label{densitymin}
e_0[\kappa,\psi]/k_{\mathrm{B}}T=\frac{1}{2}\tilde{A}(\psi)\kappa^2.
\end{eqnarray}
Substituting this equation into eq. (\ref{npdb}), one can find $p_{\mathrm{2D}}(\theta(L))$, in this limit, as follows
\begin{align}
\label{npdbmin}
p_{\mathrm{2D}}(\theta(L)&) = \mathcal{N}\int_{-\infty}^{+\infty}\frac{dK}{2\pi}e^{iK\theta}\nonumber\\
&\times\int \mathcal{D}[\kappa]\mathcal{D}[\psi]\exp\left(-\int_0^L(\frac{\tilde{A}(\psi)}{2}\kappa^2 +iK\kappa)ds\right)\nonumber\\
&=\sqrt{\frac{2P_{\mathrm{2D}}^{min}}{\pi L}}\times\exp\left(-\frac{P_{\mathrm{2D}}^{min}\theta^2}{2L}\right),
\end{align}
where
\begin{eqnarray}
\label{P2dmin_ext}
P_{\mathrm{2D}}^{min} = \frac{\int_0^{2\pi}d\psi/\tilde{A}(\psi)^{1/2}}{\int_0^{2\pi}d\psi/\tilde{A}(\psi)^{3/2}}.
\end{eqnarray}
We numerically solve this equation for any given $A_1$ and $A_2$ in the rest of this work.

From equations (\ref{npdbmax2}) and (\ref{npdbmin}), it can be deduced that the anisotropic elastic model in 2D behaves like a isotropic model with an effective persistence length, $P_{\mathrm{2D}}$, which is a function of $A_1$, $A_2$, $C$, and $\omega_0L$. As $P_{\mathrm{2D}}$ varies by the strength of twist rigidity, the upper and lower limits of $P_{\mathrm{2D}}$ are given by $P_{\mathrm{2D}}^{max}$ (eq. (\ref{P2dmax})) and $P_{\mathrm{2D}}^{min}$ (eq. (\ref{P2dmin_ext}), respectively. We perform Monte-Carlo (MC) simulations to evaluate $P_{\mathrm{2D}}$ between these two extreme limits.   
\subsection{Monte-Carlo Simulations}
To calculate the statistical properties of the chain, we performed Monte-Carlo (MC) simulations of a discrete elastic model from eq. (\ref{elasticener}). Here, the chain consists of beads were connected to adjacent beads via a link length $0.34\,\mathrm{nm}$, and without excluded volume interactions (a phantom chain). The Metropolis algorithm with appropriate Boltzmann distributions was used to construct equilibrium configurations of the chain. Our simulations were done with a linear chain containing 600 beads in 2D and 3D conformations. To estimate the statistical errors, we performed several realizations with different initial conditions.  
\begin{figure}[thb]
\begin{center}
\centering
\subfigure{
\label{fig:4a}
\includegraphics[scale=0.5]{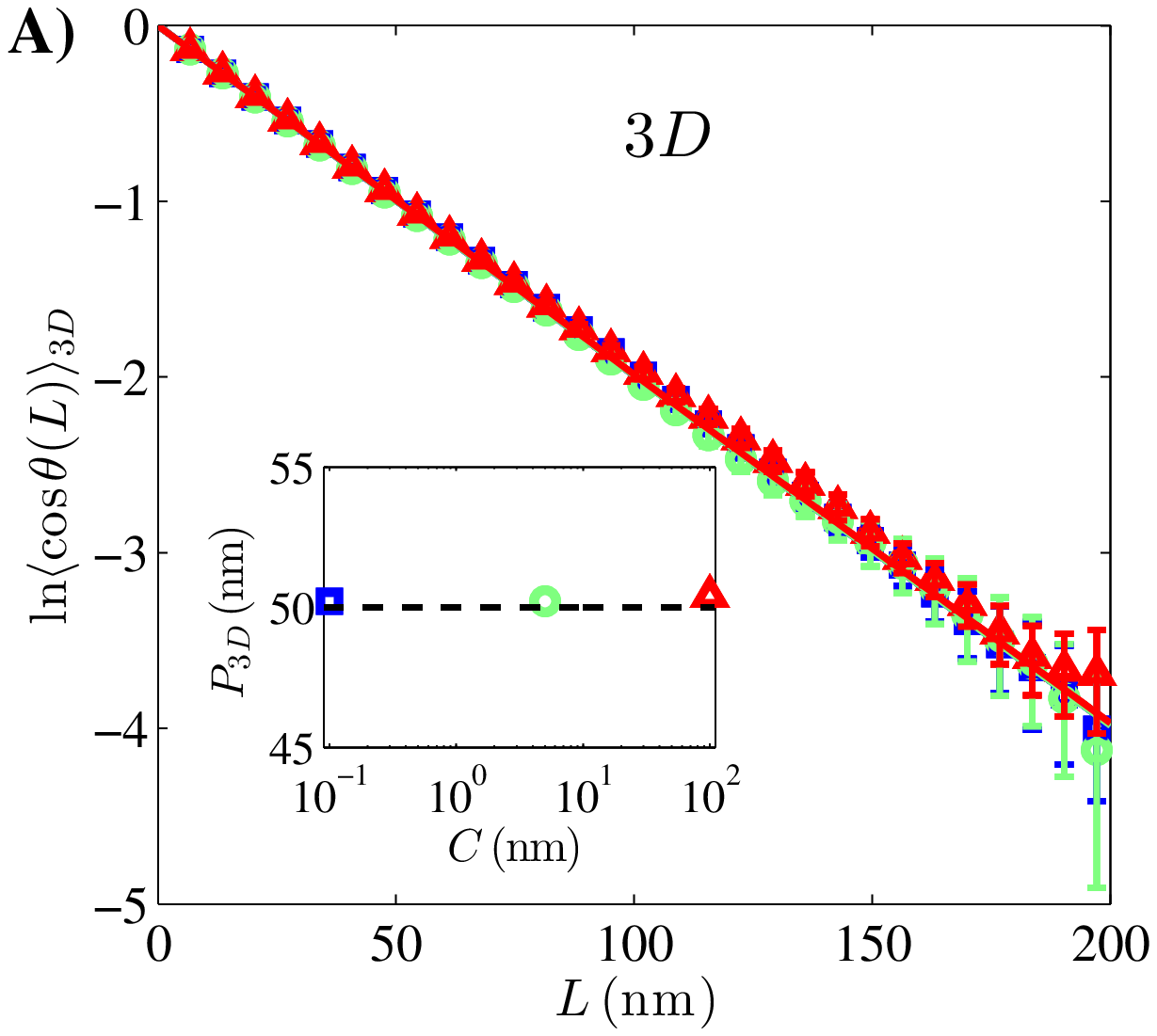}  
}\\
\subfigure{
\label{fig:4b}
\includegraphics[scale=0.5]{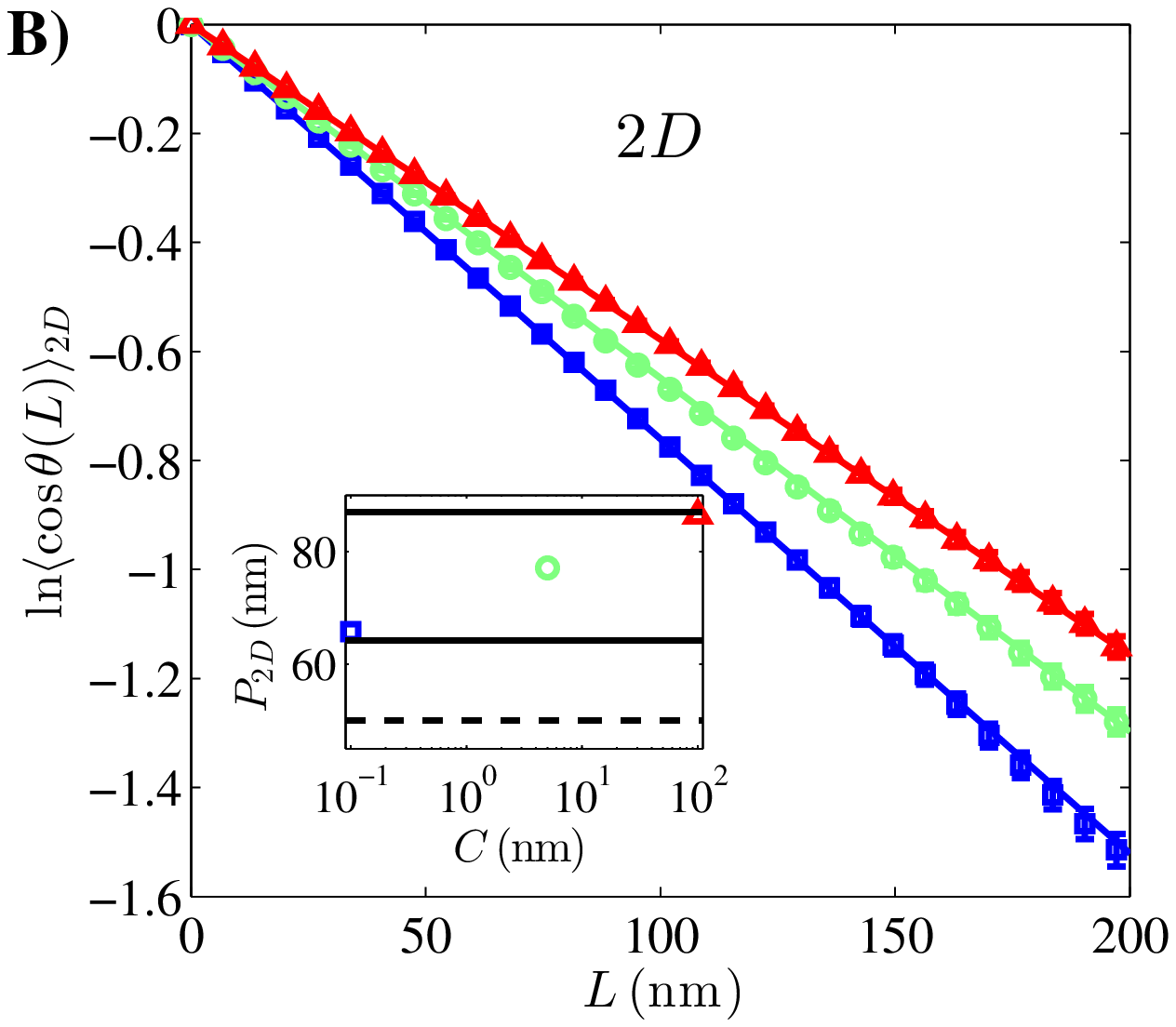}    
}
\end{center}
\caption{The MC results for the tangent-tangent correlation function in 3D (A) and 2D (B) for $A_1=275\,\mathrm{nm}$, $A_2=27.5\,\mathrm{nm}$ with a harmonic mean of $50\,\mathrm{nm}$ (dashed lines in the insets). The data points corresponding to $C=0.1$ (open, blue squares), $5$ (open, green circles), and $100\,\mathrm{nm}$ (open, red triangles), and the straight lines indicate liear fits to data. The insets show the masuared persistence lengths from the slope of the fitted lines (see text). The solid lines, in the inset of panel (B), correspond to the upper ($87.0\,\mathrm{nm}$) and lower ($64.2\,\mathrm{nm}$) limits of $P_{\mathrm{2D}}$ versus $C$ (the equations (\ref{P2dmax_ext}) and (\ref{P2dmin_ext}), respectively). \label{fig:4}}
\end{figure}
\section{results and Discussion}
\subsection{The elastic properties of the long anisotropic chain}\label{RD1}
Similar to experiments~\cite{Rivetti1996,Faas2009,Kundukad2014}, we used the tangent-tangent correlation function as well as the mean square of end-to-end distance to extract the effective persistence length of a long chain from the MC simulations. To address the effects of 2D confinement on the flexibility of chain, we compare the effective persistence length in both 2D and 3D conformations.
\begin{figure}[h]
\includegraphics[scale=0.55]{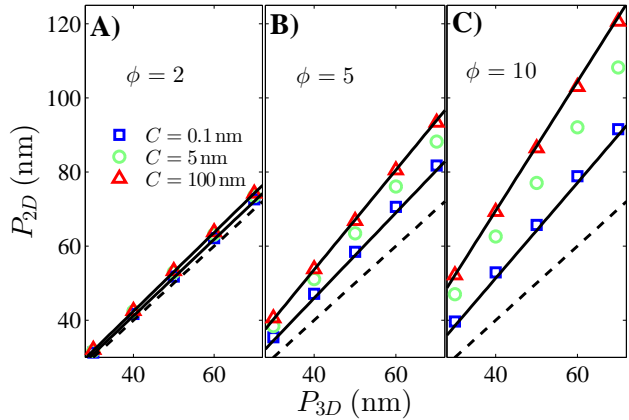}
\caption{The dependence of $P_{\mathrm{2D}}$ on $P_{\mathrm{3D}}$ for different values of $C$ and $\phi$, as indicated. Solid lines correspond to the upper and lower limits of $P_{\mathrm{2D}}$ (i.e. $P_{\mathrm{2D}}^{max}$ and $P_{\mathrm{2D}}^{min}$, respectively), and dashed lines are with slope equal to 1. Error bars (not shown) are about the size of the markers.}
\label{fig:5}
\end{figure}
\begin{figure}[thb]
\begin{center}
\centering
\subfigure{
\label{fig:6a}
\includegraphics[scale=0.39]{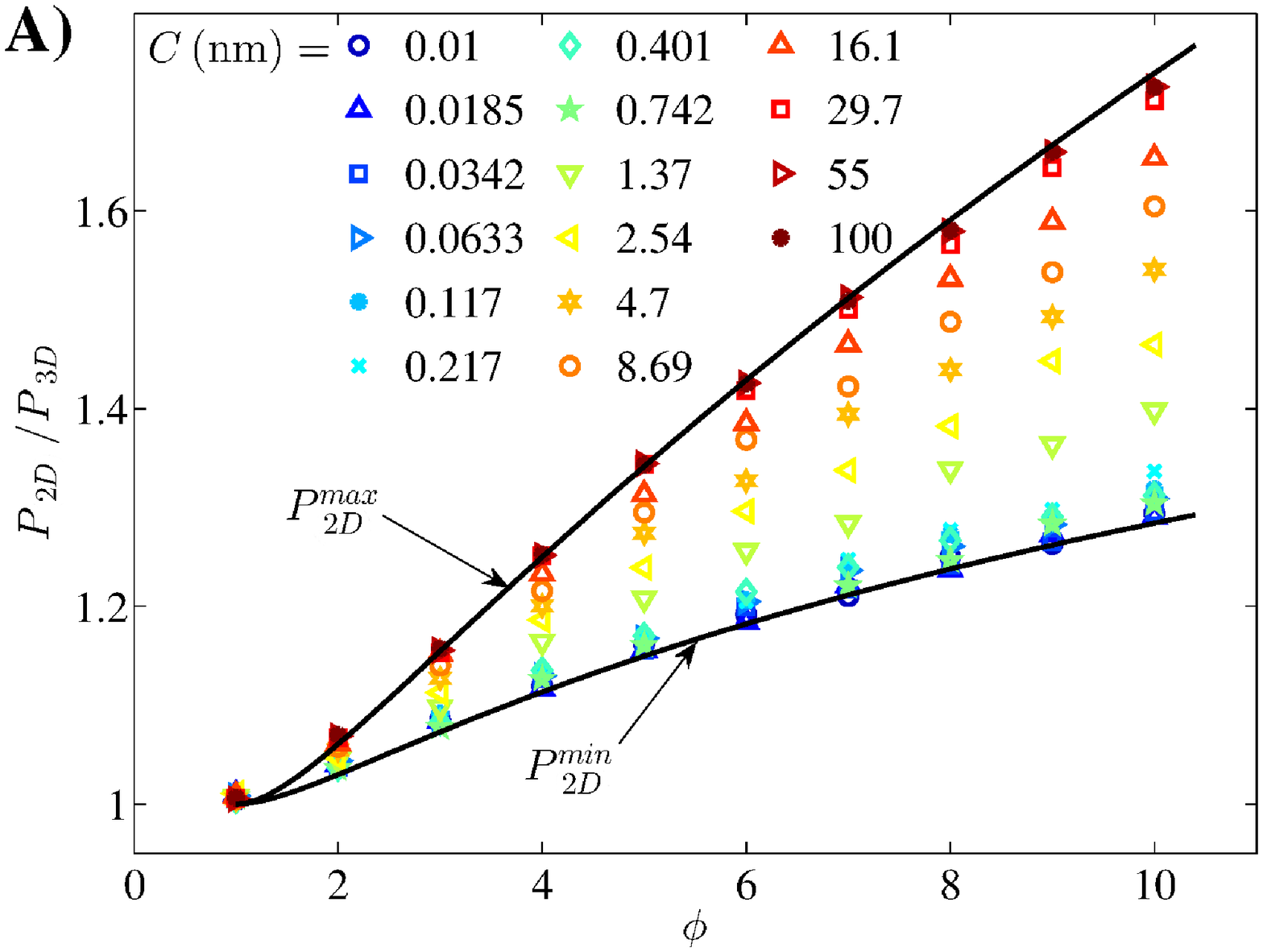}  
}\\
\subfigure{
\label{fig:6b}
\includegraphics[scale=0.4]{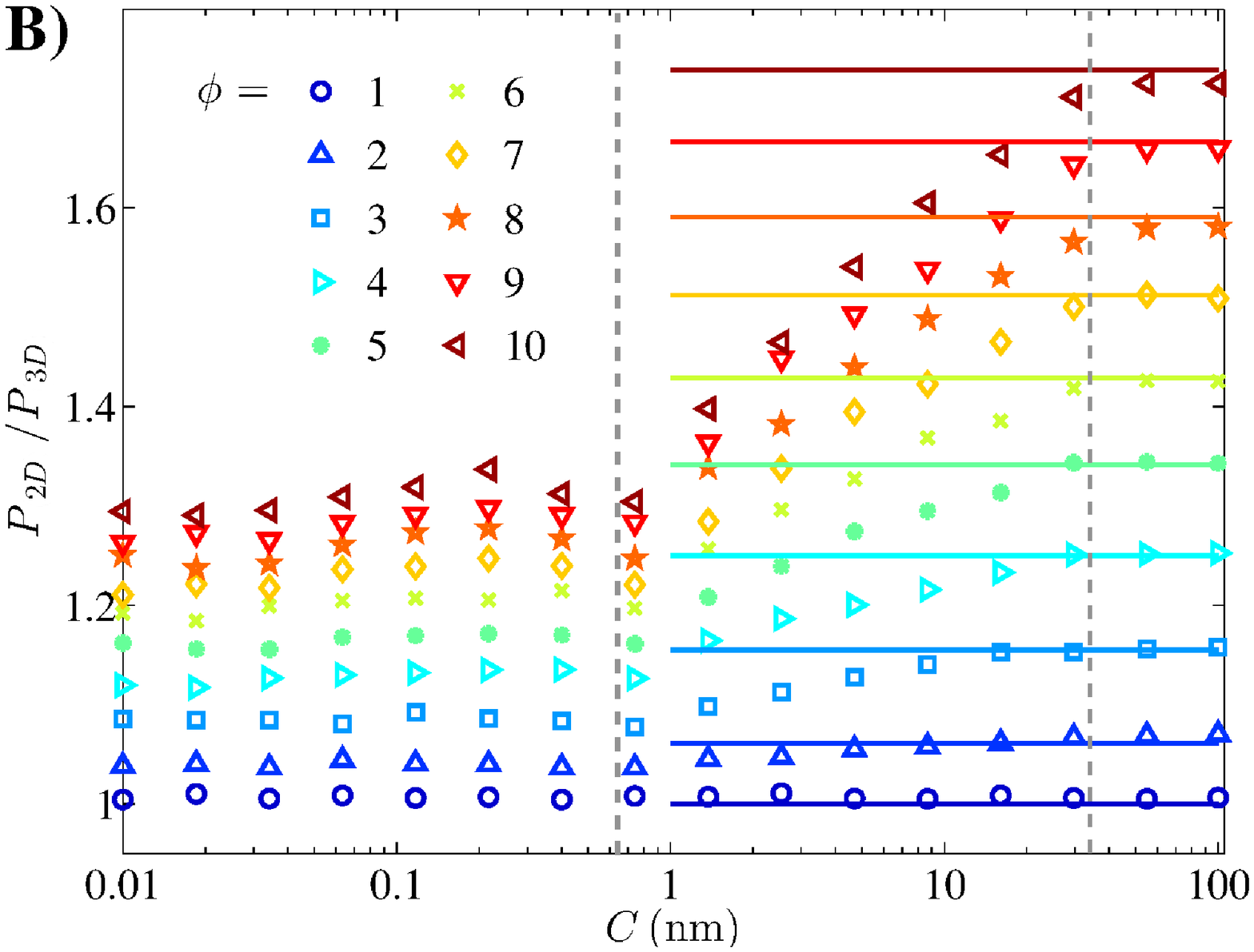}    
}
\end{center}
\caption{A) Dependence of $P_{\mathrm{2D}}/P_{\mathrm{3D}}$ on $\phi$ for different values of $C$ in the range $0.01-100\,\mathrm{nm}$ (from dark blue to dark red). The solid curves correspond to theoretical predictions for $P_{\mathrm{2D}}^{min}$ (when $C\rightarrow 0$) and $P_{\mathrm{2D}}^{max}$ (when $C\rightarrow \infty$). B) Dependence of $P_{\mathrm{2D}}/P_{\mathrm{3D}}$ on $C$ for different values of $\phi$. The horizental solid lines indicate the value of $P_{\mathrm{2D}}^{max}$ for the corresponding $\phi$, and the region between the vertical dashed (gray) lines indicates the rigion with strong twist-bend coupling regime. Error bars (not shown) are about the size of the markers. \label{fig:6}}
\end{figure}

It is well known that, the tangent-tangent correlation function of a free, long and highly twisted anisotropic model in 3D decays as $\langle\cos\theta(L)\rangle_{\mathrm{3D}} = \exp(-L/P_{\mathrm{3D}})$, where $P_{\mathrm{3D}}$ is given by the harmonic mean of the hard and soft bending rigidities, i.e. $A_1$ and $A_2$,~\cite{Eslami-Mossallam2008,Lankas02,Schurr1985}
\begin{eqnarray}
\label{P3d}
P_{\mathrm{3D}} = 2\left(1/A_1 + 1/A_2\right)^{-1}.
\end{eqnarray}
Figure \ref{fig:4}(A) compares the MC results of the 3D correlation function, $\langle\cos\theta(L)\rangle_{\mathrm{3D}}$, for the chains with $A_1=275\,\mathrm{nm}$, $A_2=27.5\,\mathrm{nm}$ (i.e. $\phi=10$) and three different values of $C$, i.e. $0.1$ (open blue squares), $5$ (open green circles), and $100\,\mathrm{nm}$ (open red triangles). $P_{\mathrm{3D}}$ is determined by the slope of the best fitted lines to $-L/\mathrm{ln}\langle\cos\theta(L)\rangle_{\mathrm{3D}}$ (the solid lines). As can be seen from the inset of fig \ref{fig:4}(A), $P_{\mathrm{3D}}$ is independent on $C$ and equal to the harmonic mean of $A_1$ and $A_2$ (i.e. $50\,\mathrm{nm}$, dashed line). Fig \ref{fig:4}(B) shows the correlation function in 2D, $\langle\cos\theta(L)\rangle_{\mathrm{2D}} = \exp(-L/2P_{\mathrm{2D}})$, here a clear dependence on $C$ is evident. We found that $P_{\mathrm{2D}}$ is always greater than $P_{\mathrm{3D}}$ (dashed line in the inset of fig \ref{fig:4}(B)) and varies from $P_{\mathrm{2D}}^{min}=64.2\,\mathrm{nm}$ for $C=0.1\,\mathrm{nm}$ to $P_{\mathrm{2D}}^{max}=87.0\,\mathrm{nm}$ for $C=100\,\mathrm{nm}$ (the solid lines in the inset of fig \ref{fig:4}(B)). Since the zero-energy configuration of a curved anisotropic chain is not planar~\cite{Norouzi2008}, then it takes energy to enforce the chain in 2D confinement. This extra energy makes the chain stiffer in 2D than 3D.   

All sets of $A_1$ and $A_2$ which results to the same $P_{\mathrm{3D}}$ but different $\phi$ values are given by
\begin{eqnarray}
\label{A1A2}
A_1 &=& P_{\mathrm{3D}}/(1-\lambda),\nonumber\\
A_2 &=& P_{\mathrm{3D}}/(1+\lambda),
\end{eqnarray}
where $\lambda = \frac{\phi-1}{\phi+1}$. Therefore, it is convenient to consider $P_{\mathrm{2D}}$ as a function of $P_{\mathrm{3D}}$, $\phi$, and $C$. Fig \ref{fig:5} shows a fairly linear relationship between $P_{\mathrm{2D}}$ and $P_{\mathrm{3D}}$ for different values of $\phi$ and $C$. We therefore expect that the ratio $P_{\mathrm{2D}}/P_{\mathrm{3D}}$ is independent of $P_{\mathrm{3D}}$. Fig \ref{fig:6} shows the dependence of this ratio on $\phi$ and $C$. As fig \ref{fig:6}(A) indicates, $P_{\mathrm{2D}}$ is always greater than $P_{\mathrm{3D}}$ when $\phi\neq1$ (i.e. anisotropic bending), and they are equal at $\phi=1$ (i.e. isotropic bending). It can be seen that $P_{\mathrm{2D}}$ as well as its lower and upper limits $P_{\mathrm{2D}}^{min}$ and $P_{\mathrm{2D}}^{max}$ increases with $\phi$. The twist-bend coupling which is reflected in the dependence of $P_{\mathrm{2D}}$ on $C$, also becomes stronger with increasing $\phi$. As fig \ref{fig:6}(B) shows, $P_{\mathrm{2D}}$ strongly depends on the twist rigidity, $C$, in the range of $1-30\,\mathrm{nm}$ (the region between two vertical dashed line), and beyond that is approximately constant and given by $P_{\mathrm{2D}}^{max}$ (eq. (\ref{P2dmax_ext})) and $P_{\mathrm{2D}}^{min}$ (eq. (\ref{P2dmin_ext})) for $C\gtrsim30\,\mathrm{nm}$ and $C\lesssim1\,\mathrm{nm}$, respectively.
\begin{figure}[h]
\includegraphics[scale=0.5]{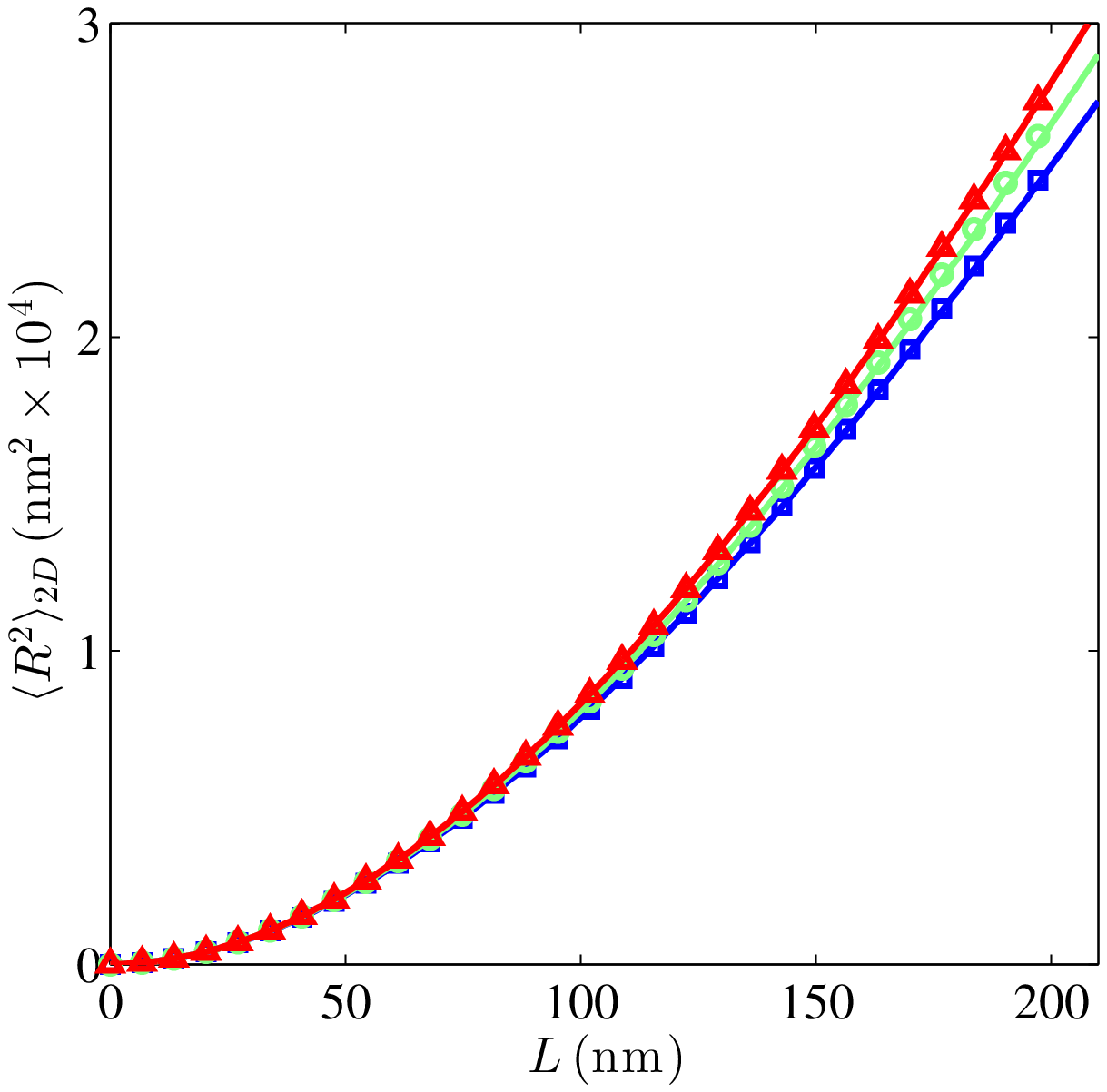}
\caption{The mean-square of end-to-end distance $\langle R^2\rangle_{\mathrm{2D}}$ versus the chain length $L$. The data points correspond to the MC simulations of chains with $P_{\mathrm{3D}}=50\,\mathrm{nm}$, $\phi=10$ and different values of $C$ (same as fig (\ref{fig:4})) and the solid curves are the predictions. Error bars (not shown) are about the size of the markers.}
\label{fig:7}
\end{figure}

Another way to calculate the persistence length is based on the mean-square of end-to-end distance $\langle R^2\rangle_{\mathrm{2D}}$. Fig \ref{fig:7} shows the MC results of $\langle R^2\rangle_{\mathrm{2D}}$ lie perfectly on the predictions of eq. (\ref{endtoend2d}), where $P_{\mathrm{2D}}$ is substituting from fig \ref{fig:4}(B). It means that the persistence length which are extracted from the mean-square of end-to-end distance are so close to those calculated from the tangent-tangent correlation function, and the relative error between these two is less than $1\%$.

\subsection{Stretching anisotropic chain in 2D}
\begin{figure}[h]
\includegraphics[scale=0.6]{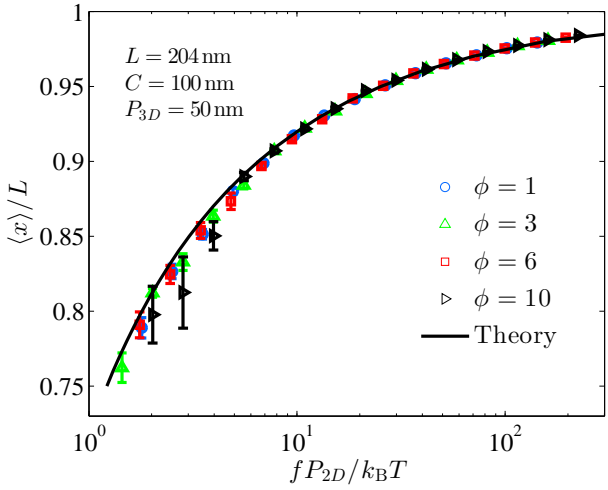}
\caption{Semi-logarithmic plot of the relative extension, $\langle x\rangle/L$, versus the scaled force, $fP_{\mathrm{2D}}/k_{\mathrm{B}}T$, for the anisotropic chains with $L=204\,\mathrm{nm}$, $C=100\,\mathrm{nm}$, $P_{\mathrm{3D}}=50\,\mathrm{nm}$, and different values of $\phi$ (as indicated) in 2D. $P_{\mathrm{2D}}$ for each set of data was extracted from fig (\ref{fig:6}(B)). Solid curve corresponds to the theoretical prediction of eq. (\ref{forceextension2d}).}
\label{fig:8}
\end{figure}
We also studied the entropic stretching of the anisotropic chain in response of an external force in 2 dimensions using MC simulations. The external potential $U=-f.x$ is added to the the elastic energy (eq. (\ref{elasticener})), where $f$ is the magnitude of the external force which is exerted on the last bead of the chain, and $x$ is the component of the end-to-end vector in the direction of the external force. It is known that the force versus extension curve of an isotropic chain (i.e. $A_1=A_2=P$) in 2D is given by~\cite{Prasad2005} 
\begin{eqnarray}
\label{forceextension2d}
fP/k_{\mathrm{B}}T = \frac{1}{16}\left(6\frac{\langle x\rangle}{L}-1+\left(1-\frac{\langle x\rangle}{L}\right)^{-2}\right),
\end{eqnarray}
where $\langle x\rangle$ is the average extension along the force direction. Fig (\ref{fig:8}) shows the 2D force-extension curve for the chains with $L=204\,\mathrm{nm}$, $C=100\,\mathrm{nm}$, $P_{\mathrm{3D}}=50\,\mathrm{nm}$, and different values of $\phi$, i.e. 1 (open blue circles), 3 (open green triangles), 6 (open red squares), and 10 (open black triangles). As it can be seen, each set of force-extension data perfectly lies on the theoretical prediction of eq. (\ref{forceextension2d}) (solid curve), when the external force is scaled by its corresponding $P_{\mathrm{2D}}$ (extracted from fig \ref{fig:6}(B)). This shows the force-extension characteristic of an anisotropic chain is same as an isotropic chain with the appropriate persistence length of $P_{\mathrm{2D}}$. 
\subsection{The elasticity of the anisotropic chain at small length-scales}
\begin{figure}[h]
\includegraphics[scale=0.5]{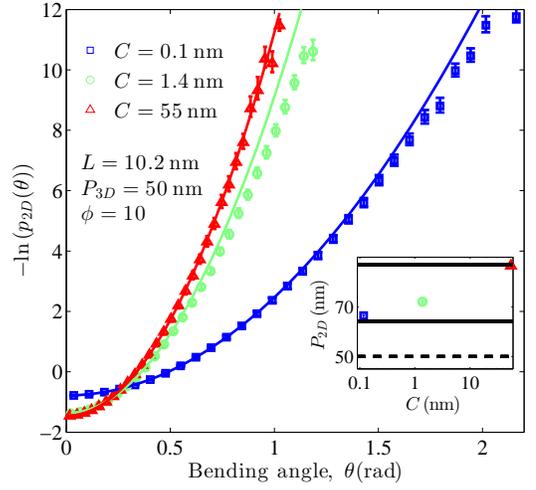}
\caption{The probability distribution $p_{\mathrm{2D}}(\theta)$ for the angle $\theta$ between tangents of two points separated by a contour length of $L=10.2\,\mathrm{nm}$. The data points correspond to MC simulations of the chains with $P_{\mathrm{3D}}=50\,\mathrm{nm}$, $\phi=10$, and three different values of $C$, as indicated. Solid curves are the best parabolic fits to the data. Inset: Semi-logarithmic plot of $P_{\mathrm{2D}}$ versus $C$. The solid (black) lines correspond to the upper (i.e. $P_{\mathrm{2D}}^{max}$) and lower (i.e. $P_{\mathrm{2D}}^{min}$) limits of $P_{\mathrm{2D}}$, and dashed line indicates $P_{\mathrm{3D}}$.}
\label{fig:9}
\end{figure}
To investigate the flexibility of the anisotropic model at short length-scales, we computed the negative logarithm of the probability distribution of bending angle, $-\mathrm{ln}\left(p_{\mathrm{2D}}(\theta(L))\right)$. Fig \ref{fig:9} shows the result for $10.2\,\mathrm{nm}$ chain with $P_{\mathrm{3D}} =50\,\mathrm{nm}$, $\phi=10$, and $C=0.1$, $1.4$, and $55\,\mathrm{nm}$. The effective persistence length of the chain at this length can be extracted by fitting a parabola of the form $(P_{\mathrm{2D}}/2L)\theta^2+constant$ to the data (see the inset of fig \ref{fig:9}), and is in good agreement with our previous result (see fig \ref{fig:4}(B)).
\begin{figure}[h]
\includegraphics[scale=0.5]{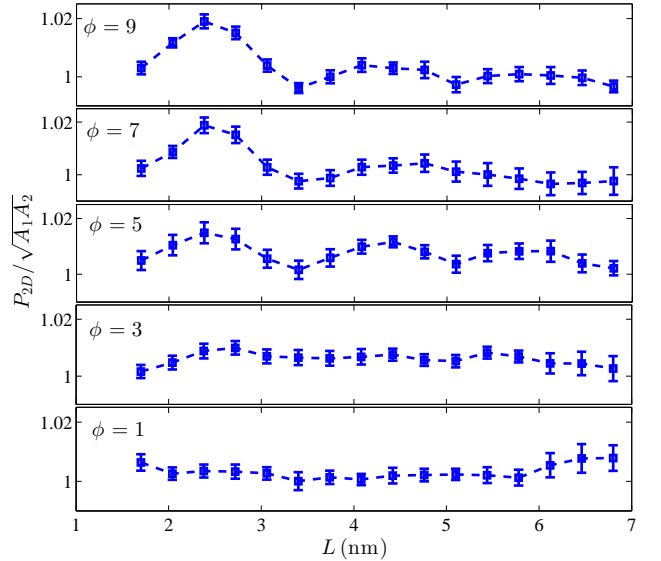}
\caption{The length dependent of the ratio of $P_{\mathrm{2D}}/\sqrt{A_1A_2}$ for $C=100\,\mathrm{nm}$, $P_{\mathrm{3D}}=50\,\mathrm{nm}$, and different values of $\phi$, as indicated. Dashed curves serve as guides for eyes.}
\label{fig:10}
\end{figure}
\begin{figure}[h]
\includegraphics[scale=0.45]{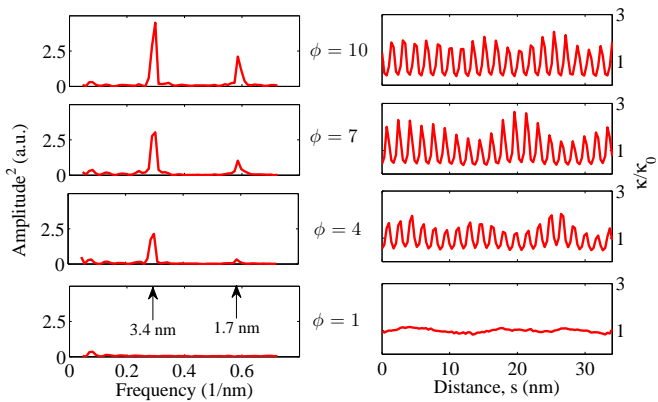}
\caption{Right: The normalized curvature, $\kappa(s)/\kappa_0$ (with $\kappa_0=2\pi/L$), of the averaged configuration of a two-dimensional anisotropic loop at room temperature. The data correspond to $L=34\,\mathrm{nm}$, $C=100\,\mathrm{nm}$, $P_{\mathrm{3D}}=50\,\mathrm{nm}$, and different values of $\phi$, as indicated. Left: Fourier spectra of the curvature profile. The arrows indicate the two main periodic components of the curvature.}
\label{fig:11}
\end{figure}

Due to the intrinsic helicity and bending anisotropy of the DNA molecule, we expect that the effective persistence length at small length scales oscillates with a period of $\pi/\omega_0\backsimeq1.7\,\mathrm{nm}$. Using $-\mathrm{ln}\left(p_{\mathrm{2D}}(\theta(L))\right)$ we calculate $P_{\mathrm{2D}}$ for the segment lengths between $1.7$ to $6.8\,\mathrm{nm}$. Fig \ref{fig:10} compares the ratio $P_{\mathrm{2D}}/\sqrt{A_1A_2}$ for $C=100\,\mathrm{nm}$ and different value of $\phi=1$, $3$, $5$, $7$, and $9$. It can be seen that $P_{\mathrm{2D}}$ oscillates with a period of about $1.7\,\mathrm{nm}$ and decays to its extreme value of $\sqrt{A_1A_2}=P_{\mathrm{3D}}/\sqrt{1-\lambda^2}$ by increasing the length. This oscillation is amplified if the strength of bending anisotropy, $\phi$, increases. 

The oscillations are due to the formation of kinks with periodic arrangement in two-dimensional ground-state conformation of a bent and twisted anisotropic chain~\cite{Mohammad-Rafiee2005,Bijani2006}. We performed MC simulations for a loop with $L=34\,\mathrm{nm}$, $C=100\,\mathrm{nm}$, $P_{\mathrm{3D}}=50\,\mathrm{nm}$, and different values of $\phi=1$, $4$, $7$, and $10$. As the right column of fig \ref{fig:11} shows, the curvature along the loop is not uniform and it is localized with a periodic arrangement (which leads to the kink formation). The Fourier spectrum of the curvature reveals two main periodic components, with helical ($2\pi/\omega_0\simeq3.4\,\mathrm{nm}$) and half helical ($\pi/\omega_0\simeq1.7\,\mathrm{nm}$) periods (the two arrows in fig \ref{fig:11}). This half helical-pitch periodicity is a result of the anisotropic model and vanishes at isotropic model (i.e. $\phi=1$). The amplitude of this component increases by increasing the strength of anisotropy, $\phi$.  
\subsection{Estimation of the anisotropic bending of dsDNA}
Sequence dependence and bending anisotropy of dsDNA has been widely noticed in base-pair steps
approaches, by partitioning the DNA deformation energy through six local variables,
slide, shift, rise, tilt, roll and twist~\cite{Mergell}. The rigidity parameters corresponding to these six variables are extracted from their standard deviation~\cite{Olson1998}. Therefore, the ratio of bending rigidities, $\phi$, can be determined by $\phi\sim(\sigma_{roll}/\sigma_{tilt})^2$, where $\sigma_{roll}$ and $\sigma_{tilt}$ denote the standard deviations of bending angles in soft (roll) and hard (tilt) directions, respectively~\cite{Olson1993}. A survey of the values of $\phi$ obtained by different techniques is presented in Table \ref{table:2}. Despite the diversity, the value of $\phi$ lies in the range of $2-6$.
\begin{table}[ht]
\caption{Some reported values of $\phi$ measured by different techniques.\label{table:2}}
\begin{tabular}{p{4.5cm}p{1.25cm}p{1.75cm}c}
\hline
\hline Investigators & $\phi$ &  Method & Ref. \\
\hline 
Zhurkin et al. (1991) & $2-4$ &  MC\footnote{Monte-Carlo simulations} & \cite{Zhurkin1991} \\
Olson et al. (1998) & $1-5$ & XRC\footnote{X-ray crystallography of protein-DNA complexes}  & \cite{Olson1998} \\
El Hassan \& Calladine (1997) & $\sim4$ & XRC &  \cite{ElHassan1997}\\
Richmond et al (2003) & $\sim2$ & XRC & \cite{Richmond} \\
Chua et al. (2012) & $2-4$ &XRC &  \cite{Chua2012} \\
Stefl et al. (2004) & $\sim3$ & NMR\footnote{Nuclear magnetic resonance spectroscopy} & \cite{Stefl2004} \\
Dornberger et al. (1998) & $\sim2$ & NMR & \cite{Dornberger1998}\\
Lankas et al. (2000)& $\sim2$ & MD\footnote{All-atom Molecular Dynamic simulations} & \cite{Lankas2000} \\
Lankas et al. (2003) & $\sim2$ & MD & \cite{Lankas2003} \\
Lankas et al. (2009) & $\sim3$ & MD & \cite{Lankas2009} \\
Lankas et al. (2010) & $2-3$ & MD \& NMR &  \cite{Lankas2010} \\
Bishop (2005) & $1.56$ & MD & \cite{Bishop2005} \\
Lavery et al. (2009) & $\sim2.5$ & MD & \cite{Lavery2009} \\
Perez et al. (2005) & $\sim2$ & MD & \cite{Perez2005} \\
Perez et al. (2008) & $\sim2$ & MD & \cite{Perez2008} \\
Becker \& Everaers (2007)& $1.8$ & MD & \cite{Becker2007} \\
Teng \& Hwang (2015) & $2-6$ & MD & \cite{Teng2015} \\ 
Balasubramanian et al. (2009) & $\sim2$ & NAD\footnote{Nucleic Acid Database~\cite{Berman1992}} & \cite{Balasubramanian2009}\\
\hline 
\end{tabular}
\end{table}

For B-DNA the twist rigidity $C$ is in the range of $40-110\,\mathrm{nm}$~\cite{Neukirch2004,Lipfert2010a}. Therefore, to estimate the anisotropy strength, $\phi$, we can use the relation of $P_{\mathrm{2D}}/P_{\mathrm{3D}}=(\phi+1)/\sqrt{4\phi}$, which is valid for $C\gtrsim30\,\mathrm{nm}$. Assuming the average values for $P_{\mathrm{2D}}$ and $P_{\mathrm{3D}}$ are given by $55.8 \pm 3.5$ and $43.3 \pm 3.7 \, \mathrm{nm}$, respectively (see fig \ref{fig:1}), we obtain $\phi=4.4\pm1.8$, which falls in the expected range (see Table \ref{table:2}).
\section{conclusions}
In summery, we have shown that enforcing the chain into a two dimensional conformation increases its stiffness. Our analytical approach and MC simulations showed that due to a twist-bend coupling in the 2D anisotropic model, the effective persistence length depends on the twist rigidity, and reaches soon to its maximum value when $C\gtrsim30\,\mathrm{nm}$. In this limit, the 2D persistence length is given by the geometric mean of the hard and soft bending rigidities, instead of the harmonic mean in 3D. In addition, we show that the twist-bend coupling leads to the formation of kinks along a curved chain as previously predicted using the energy minimization treatment~\cite{Mohammad-Rafiee2005,Bijani2006}.

We estimated the bending anisotropy of dsDNA, and it turns out that the hard bending rigidity is almost 4 times larger than the soft bending rigidity. this is compatible with the estimates in the literature, although we expect that this value is sensitive to the experimental conditions. Our analytical procedure can be used as a way to estimate the bending rigidities of other anisotropic bending polymers, such as nano-ribbons and dsRNA.

\bibliography{library}
\end{document}